\documentclass[a4paper,12pt,nofootinbib]{revtex4-1}

\usepackage{amsmath,amssymb}
\usepackage{braket}
\usepackage[dvipsnames]{xcolor}
\usepackage{graphicx}
\usepackage{upgreek}

\newcommand{\rd}{\mathrm{d}}

\begin{document} 

\title{Non-perturbative aspects of galileon duality}

\author{Peter Millington}
\email{p.millington@nottingham.ac.uk}
\author{Florian Niedermann}
\email{florian.niedermann@nottingham.ac.uk}
\author{Antonio Padilla}
\email{antonio.padilla@nottingham.ac.uk}

\affiliation{School of Physics and Astronomy, University of Nottingham,\\ Nottingham NG7 2RD, United Kingdom}

\date{July 6, 2018}

\begin{abstract}
We study quantum aspects of the galileon duality, especially in the case of a particular interacting galileon theory that is said to be dual to a free theory through the action of a simultaneous field and coordinate transformation. This would appear to map a theory with multiple vacua to one with a unique vacuum state. However, by regulating the duality transformation using external sources, we are able to preserve the full vacuum structure in the dual frame. By explicitly calculating the one-particle irreducible effective action on a  maximally symmetric background, we identify a semi-classical contribution to the Wightman functions that has not been taken into account in previous analyses due to the singular  point in the duality map. This  may affect its spectral properties at high energy scales. These observations cast doubt on the main evidence in support of a non-local UV structure for galileons. 
\end{abstract}

\maketitle

\section{Introduction}

The accelerated expansion of the Universe is now well established by a slew of observational data (see e.g.~Refs.~\cite{planck2015,SDSS2014}), yet the exact origin of this acceleration remains an open question (for a review, see~Ref.~\cite{DEreview}). The ``simplest'' explanation, in which dark energy corresponds to a very low-scale cosmological constant, is severely challenged by naturalness considerations~\cite{wein, pol,Cliff, CCreview}, leading  many phenomenologists to explore alternative mechanisms including  infrared modifications of General Relativity operative on cosmological scales~\cite{myreview, JKreview}. A special class of derivatively coupled scalar theories have played a significant role in these explorations. These are the \emph{galileons}~\cite{Fairlie1,Fairlie2,Nicolis2009,bigal1,Mgal, kurtmgal,bigal2}.

Galileons arise in  a specific limit of the DGP braneworld
model~\cite{Dvali2000}, where all gravitational degrees of freedom
apart from the helicity-zero mode are decoupled. It can also be shown that the same structure emerges
  from the decoupling limit of massive gravity~\cite{Rham2010}.  Up to a total derivative, the galileon  Lagrangians are invariant under the galileon
symmetry $\pi \to \pi +c + v_\mu x^\mu$, a fact that gives rise to enhanced soft limits for scattering amplitudes~\cite{trnka1,trnka2, toby}. The most important feature of galileon theories, however,  is the presence of derivative interactions that allow for Vainshtein screening~\cite{Vainshtein1972b,vainintro,dyson}.  When a phenomenological model deviates significantly from General Relativity on cosmological scales,  compatibility with observational constraints at shorter distances~\cite{Will} can be achieved thanks to the Vainshtein mechanism,  exploiting derivative non-linearities to screen fifth forces in the vicinity of massive sources like the
Earth or the Sun. The potential importance of screening mechanisms in developing viable models of \emph{self-tuning} was recently highlighted in~Ref.~\cite{nogo}.

Despite these interesting properties, the theoretical and phenomenological consistency of galileon theories remains an open issue. In particular, the structure of the galileon interactions would indicate a breakdown of perturbative unitarity, often at an unacceptably low scale. For example, for a dRGT graviton whose mass is around the Hubble scale $H_0$, perturbative unitarity breaks down at the  scale \smash{$\Lambda_3 \sim (M_{Pl}H_0^2)^{1/3} \sim \left( 1000 \, \mathrm{km}\right)^{-1}$}~\cite{Rham2010,clare} (see, however, Ref.~\cite{scaleup}, for an interesting recent development).  Standard effective field theory (EFT) methods require us to cut the theory off,  eliminating its predictive power on observationally relevant scales above this cut off. Worse still, EFT corrections, arising from integrating out whatever new physics is introduced  to preserve perturbative unitarity, can  contaminate the dynamics of  Vainshtein screening even out to larger distances~\cite{Kaloper2015a}.  These considerations make it imperative that we better understand the UV completion of galileon theories before taking them seriously as a candidate model of nature to be probed by future experiments.

Doubts have been raised as to whether or not galileon theories can admit a UV completion in the standard Wilsonian sense~\cite{Adams2006}. However, such conclusions, based on S-matrix analyticity, are a little premature. By gently deforming the galileon theory in the far infra-red (in a technically natural way), one can easily evade the analyticity no go results. For example, we can give the galileon a small mass~\cite{massivegal} and/or introduce a tiny galileon-breaking interaction of the form $g (\partial \pi)^4$~\cite{amppos}. Nevertheless, there is no known Wilsonian UV completion of galileons, fuelling the perception that no such completion exists and encouraging alternative ideas.  

No doubt motivated by these considerations, Keltner and Tolley~\cite{Keltner2015} have recently argued that galileons proceed towards a
non-standard, {i.e.}\ non-Wilsonian UV completion, in which there is a certain degree of non-locality at high
energies. This unconventional UV picture is inferred from the
high-energy behaviour of the momentum-space Wightman functions, which
are argued to exhibit exponential growth (rather than
polynomial boundedness) above the strong-coupling scale.   To support this claim, the authors perform explicit calculations in the case of one particular galileon theory, characterized by a remarkable property: it can be mapped to a
\textit{free theory} by transforming the field and the coordinates
simultaneously~\cite{DeRham2014a,DeRham2014b}. This
``interacting-to-free'' duality is then the key to accessing the a
priori unknown and strongly coupled UV sector of the interacting
theory. The result is then used to support the notion that UV completion of galileons should now follow along the lines
of the classicalization proposal~\cite{Dvali2011a}.
  
In this paper we take a closer look at galileon duality at the quantum level,  and in particular the duality map between interacting and free galileon theories. Our interest lies in the implications for the UV sector of galileon theories and the conclusions we can draw regarding their  spectral properties beyond the perturbative level.
Classically, the interacting theory is known to exhibit Vainshtein screening when minimally coupled to a point source and  the only way to retain this behaviour in the dual free theory is via non-minimal coupling to sources. The importance of keeping track of how source couplings change via the duality map was  emphasized already in Ref.~\cite{Creminelli2015}, with superluminal propagation around non-trivial backgrounds shown to persist in the ``free''  theory on account of the non-localities in the mapped source term. At the quantum level, we note that external sources are also used to support off-shell configurations when computing the  quantum-corrected effective actions. The details of the external-source couplings are clearly going to be important in carefully maintaining the equivalence of the two theories on either side of the duality map beyond tree level.  To naively compare the two, as in Ref.~\cite{Keltner2015},  is to compare apples with oranges.

The starting point in our analysis is to examine the classical vacuum structure of the two theories, where we are presented with an immediate puzzle.  Galileons are known to support ``maximally symmetric'' field configurations of the form $\pi_\kappa(x) \propto \kappa\, x_\mu x^\mu$.  A generic galileon theory in $d$ dimensions will admit up to $d$ vacua. Clearly, a free theory has a unique trivial vacuum. Is the same true for its interacting dual? As it turns out, the interacting dual theory actually has two vacua: $\pi_0 = 0$ and $ \pi_1 \propto x_\mu x^\mu $.  The latter falls under the class of so-called self-accelerating vacua, which may have interesting cosmological applications~\cite{Nicolis2009}. The mismatch can be understood by studying the action of the duality map on the space of maximally symmetric configurations parametrized by $\kappa$. This behaves like a simple stereographic projection, leaving $\pi_0$ invariant but mapping $\pi_1$ to infinity. In other
words, $\pi_1$ corresponds to the \textit{projection point} and
therefore lacks a dual partner within the physical configuration
space. Already, this simple observation raises the suspicion that the
triviality of the dual theory is an artefact of the singular nature of
the map. The possibility of singular points in the duality map were mentioned in Ref.~\cite{DeRham2014a}, although their implications on the
  non-perturbative sector were not  explored.

The bulk of our analysis  is devoted to the calculation of the one-particle irreducible (1PI)
effective action~\cite{Jackiw1974}, which is achieved by performing a saddle-point
evaluation about the maximally symmetric ``kappa configurations''. This explicitly  illustrates the non-triviality of the theory's
vacuum structure, which is argued to affect the high-energy behaviour
of the spectral density (or, equivalently, the Wightman functions). Specifically, we show that the one-loop effective action in $d$
dimensions exhibits a violation of convexity when evaluated by expanding around configurations with  $\kappa > 1/d$,
signalling the presence of instabilities at  energy
densities close to the strong coupling scale. We also show
that the same expression for the effective action can be obtained
by employing the duality transformation in the presence of an external
source. In that case, the non-triviality is maintained in the source-dependent term in the action.

We finally consider the calculation of the Wightman functions and ask what can be reliably inferred about the high-energy behaviour of the
spectral density. By appropriately taking into account the non-trivial
vacuum structure, which again is achieved by including an external
source, we isolate an additional saddle-point contribution,
corresponding to $\pi_1$, which has been missed before. Most
importantly, its explicit cut-off dependence casts doubt on any
inference that can be drawn about the theory's UV sector from the
corresponding expression for the Wightman functions. Unfortunately, we conclude that galileon duality is not a particularly useful tool for inferring the UV behaviour of galileon theories.

%%%

\section{The galileon duality}

We begin by reviewing the details of the galileon duality, as first described in Ref.~\cite{DeRham2014a}, in the context of a particular interacting galileon theory with classical action
\begin{equation}
\label{eq:pi_action}
  S[\pi]\ =\ -\,\frac{1}{2}\int\!{\rm d}^dx\;\mathrm{det}\big(1+\Pi(x)\big)\big(\partial \pi(x)\big)^2\;,
\end{equation}
where
$
  \Pi(x) \equiv \Pi_{\mu}^{\phantom{\mu}\mu}(x)$ and $\Pi_{\mu\nu}(x) = \frac{1}{\Lambda^{\sigma}}\,\partial_{\mu}\partial_{\nu}\pi(x)
$. Unless otherwise stated, we work in $d${-}dimensional Minkowski spacetime. We employ the signature convention $(-,+,+,+)$, and $\partial_{\mu}\equiv\partial/\partial x^{\mu}$ denotes the partial derivative with respect to the spacetime coordinate $x^{\mu}$.

We can define a dual galileon $\rho$ via the following field-dependent coordinate transformation~\cite{DeRham2014a}:
\begin{subequations}
  \label{eq:duality}
  \begin{align}
    \tilde{x}^{\mu}\ &=\ x^{\mu}\:+\:\frac{1}{\Lambda^{\sigma}}\,\partial^{\mu}\pi(x)\;,\\
    x^{\mu}\ &=\ \tilde{x}^{\mu}\:-\:\frac{1}{\Lambda^{\sigma}}\,\tilde{\partial}^{\mu}\rho(\tilde{x})\;,
  \end{align}
\end{subequations}
where $\tilde{\partial}_{\mu}\equiv \partial/\partial \tilde{x}^{\mu}$
is the derivative with respect to the transformed spacetime
coordinate $\tilde{x}^{\mu}$ and $\sigma=(d+2)/2$. The volume measures transform as follows:
\begin{subequations}
  \begin{align}
    \int\!{\rm d}^dx\;\mathrm{det}\big(1+\Pi(x)\big)\ =\ \int\!{\rm d}^d\tilde{x}\;,\\
    \int\!{\rm d}^dx\ =\ \int\!{\rm d}^d\tilde{x}\;\mathrm{det}\big(1-\Sigma(\tilde{x})\big)\;,
  \end{align}
\end{subequations}
where
$
  \Sigma(\tilde{x}) \equiv \Sigma_{\mu}^{\phantom{\mu}\mu}(\tilde{x})$ and $ \Sigma_{\mu\nu}(\tilde{x}) = \frac{1}{\Lambda^{\sigma}}\,\tilde{\partial}_{\mu}\tilde{\partial}_{\nu}\rho(\tilde{x})
$.
Applying this transformation, the classical action in Eq.~\eqref{eq:pi_action}
becomes
\begin{equation}
\label{eq:rho_action}
  S[\pi]\ \equiv\ \tilde{S}[\rho]\ = \ -\,\frac{1}{2}\int\!{\rm d}^d\tilde{x}\;\big(\tilde{\partial} \rho(\tilde{x})\big)^2\;,
\end{equation}
and we see that the classical galileon duality
is extreme for this particular theory: we map a seemingly non-trivial theory
[Eq.~\eqref{eq:pi_action}] to a free one
[Eq.~\eqref{eq:rho_action}]. Whilst the equation of motion in the original $\pi$ frame is
\begin{equation}
  \label{eq:eom_pi}
  \frac{\delta S[\pi]}{\delta \pi}\ =\ \partial_\mu \left\{\mathrm{det}\big(1+\Pi\big)\partial^\mu \,\pi\right\}\:-\:\frac{1}{2\Lambda^{\sigma}}\,\partial^{\mu}\partial_{\nu}\Big\{\mathrm{det}\big(1+\Pi\big)\,\big[\big(1+\Pi\big)^{-1}\big]_{\mu}^{\phantom{\mu}\nu}\big(\partial\pi\big)^2\Big\}\,,
\end{equation}
that of the $\rho$-frame theory is simply
\begin{equation}
\frac{\delta \tilde{S}[\rho]}{\delta \rho}\ =\ \tilde{\Box}\,\rho\ =\ 0\;.
\end{equation}
The former of these equations
admits a
one-parameter family of maximally symmetric configurations of the form
\begin{equation}
\pi_{\kappa}(x)\ =\ -\:\frac{\kappa}{2}\;\Lambda^{\sigma}\,x_{\mu}x^{\mu}\;,
\end{equation}
with 
\begin{equation}
\label{eq:firstvar}
\kappa \, (1-\kappa)^{d-1}\ =\ 0 \;.
\end{equation}
We therefore have two ``kappa vacua'' for all $d>1$: $\kappa=0$ and
$\kappa=1$. These solutions are particularly interesting in that they
correspond to maximally symmetric spacetime geometries when the
galileon $\pi$ is coupled to the trace of the energy-momentum tensor
(and its back-reaction can be neglected), cf.~the discussion in Ref.~\cite{Nicolis2009}. However, it is clear that the $\rho$-frame equation of motion is only able to support the trivial solution $\rho=0$, and the $\kappa=1$ vacuum state has been lost.

In order to see how vacuum solutions may have been lost, it is instructive to consider the relationship between the dual galileons themselves~\cite{DeRham2014a}:
\begin{align}
  \label{eq:4}
\rho[\pi](\tilde{x})\ =\ \pi (\tilde{x} - \tilde{\partial}  \rho[\pi](\tilde{x}) / \Lambda^\sigma)\:+\:\frac{1}{2 \Lambda^\sigma} \, \tilde{\partial}  \rho[\pi](\tilde{x}) \cdot \tilde{\partial} \rho[\pi](\tilde{x}) \;.
\end{align}
Taking the functional $\rho$ derivative of Eq.~\eqref{eq:4} and making use of the fact that 
\begin{equation}
\label{eq:pirhoident}
\partial_{\mu}\,\pi(x)\ =\ \tilde \partial_{\mu}\,\rho(\tilde x)\;,
\end{equation}
we find
\begin{align}
  \label{eq:5}
\frac{\delta  \rho[\pi](\tilde{y})}{\delta \pi(x)} \ &= \delta( \tilde{y} - \tilde{\partial}  \rho[\pi](\tilde{y}) / \Lambda^\sigma - x)\:-\: \frac{1}{\Lambda^\sigma}\bigg[\partial \pi(y) \cdot \frac{\delta \tilde{\partial}  \rho[\pi](\tilde{y})}{\delta \pi(x)}\: -\: \tilde{\partial}  \rho[\pi](\tilde{y}) \cdot \frac{\delta \tilde{\partial}  \rho[\pi](\tilde{y})}{\delta \pi(x)}\bigg]\nonumber\\
&= \delta( \tilde{y} - \tilde{\partial}  \rho[\pi](\tilde{y}) / \Lambda^\sigma - x)\;.
\end{align}
It is important to keep in mind when performing the functional derivatives that $x$ and $\tilde y$ are variables of integration and therefore do not posses implicit dependence on either $\rho$
 or $\pi$. Using the functional chain rule, we can now write
 \begin{equation}
 \frac{\delta S[\pi]}{\delta \pi(x)}\ =\ \int\!{\rm d}^d\tilde{y}\;\frac{\delta S[\pi]}{\delta \rho(\tilde{y})}\,\frac{\delta \rho[\pi](\tilde{y})}{\delta \pi(x)}\ =\ \int\!{\rm d}^d\tilde{y}\;\delta( \tilde{y} - \tilde{\partial}  \rho[\pi](\tilde{y}) / \Lambda^\sigma - x)\,\frac{\delta \tilde{S}[\rho]}{\delta \rho(\tilde{y})}\;,
 \end{equation}
from which it is clear that the vanishing of $\delta S[\pi]/\delta \pi(x)$ does not imply the vanishing of $\delta \tilde{S}[\rho]/\delta \rho(\tilde{y})$, except when $\tilde{\partial}_{\mu}\,\rho=0$ (when tilded and untilded coordinates coincide).

The focus of the remainder of this article will be to study the behaviour of these distinct vacua under the galileon duality in more detail. Ultimately, however, we will be led to conclude that the naive application of the galileon duality can dramatically alter the vacuum structure of the theory.

%%%

\subsection{The action polynomial and vacuum diagnostics}
\label{sec:actionp}

In order to understand better the impact of the galileon duality on the kappa configurations, it is instructive to consider the \emph{action polynomial} $U(\kappa)$. The action polynomial is readily obtained by Wick rotating to Euclidean signature and evaluating the Euclidean action for Eq.~\eqref{eq:pi_action}  on $O(d)$-symmetric configurations $\pi(r)$, where \smash{$r=\sqrt{\tau^2+{\bf x}^2}$} and $\tau=it$ is Euclidean time. Provided we include the boundary terms required for a well-defined variational principle under Dirichlet boundary conditions~\cite{Padilla2012}, the result can be written explicitly in first-order form as
\begin{equation}
S_\mathrm{E}[\pi]\ =\ \frac{\Omega_{d-1}}{d^2} \int \rd r\; r^{d+1}  U(\psi)\;,
\end{equation}
where $\Omega_{d-1}$ is the volume of the unit $(d-1)$ sphere,  $\psi \equiv -\pi' / (r \Lambda^\sigma)$ and 
\begin{equation}
  U(\psi)\ =\ \frac{d}{\left( d+1 \right)} \, \Lambda^{2 \sigma} \, \left[1 - (1 +d \, \psi) (1-\psi)^{d} \right] \,.
\end{equation}
On the kappa configurations, in  Euclidean signature, we have $\pi_\kappa(r)=-\,\kappa\,\Lambda^\sigma r^2/2$, upon which $\psi$ evaluates to $\kappa$, and we recover the action polynomial $U(\kappa)$. As demonstrated explicitly in Ref.~\cite{bigal1}, this polynomial provides a very simple way to classify vacua. For example,  comparing with Eq.~\eqref{eq:firstvar}, we see that the first derivative of $U(\kappa)$ is proportional to the equation of motion evaluated
at $\pi_\kappa$:
\begin{align}
  \label{eq:17}
  \frac{\delta}{\delta \pi_x} \left( S[\pi] +\textrm{boundary terms}\right)\Big|_{\pi\,=\,\pi_\kappa}
  \ &=\ -\,d\,\kappa(1-\kappa)^{d-1}\Lambda^{\sigma} \nonumber \\
  &=\ \frac{1}{d \, \Lambda^\sigma } \, \frac{\rd  U(\kappa)}{\rd \kappa} \;.
\end{align}
It follows that the stationary points of~$ U(\kappa) $ correspond to
solutions of the classical equations of motion.  Similarly, the second derivative of $U(\kappa)$ is related to the second variation of the action:
\begin{align} \label{eq:2nd_var}
  \frac{\delta^2}{\delta \pi_x\,\delta\pi_y} \left( S[\pi] +\textrm{boundary terms} \right)\Big|_{\pi\,=\,\pi_{\kappa}}\
  \ &=\  (1-\kappa)^{d-2}(1-d\kappa)\,\Box\,\delta^d(x-y) \nonumber \\
  &=\ \left(\frac{1}{d \, \Lambda^\sigma }\right)^2 \, \frac{\rd^2  U(\kappa)}{\rd \kappa^2}\,\Box\,\delta^d(x-y) \;,
\end{align}
and its sign determines whether fluctuations about the
respective classical configuration exhibit a ghost instability
(negative sign), are strongly coupled (zero) or stable (positive
sign). The second variation evaluates to
$\Box \, \delta^4(x-y)$ for $\kappa=0$, but  it vanishes for
$\kappa=1$ and $d>2$, signalling a strongly-coupled point. For
$d\geq2$ and $\kappa=1$, the first non-vanishing variation of the
action then arises at $d$-th order.

Whilst $U(\kappa)$ can be used as a tool to diagnose ghost
instabilities (and violations of the convexity of the effective
action, as we will see later), it is not directly related to the
energy of a given configuration. Instead, the total energy $E_\kappa$ of the Lorentzian kappa configurations (within a $(d-2)$-sphere of radius $\mathfrak{R}$)
is given by
\begin{align}
  \label{eq:14}
\frac{E_{\kappa}}{\mathcal{V}_{d-2} \, \Lambda^{2\sigma}}  =  \frac{1}{2} \, \kappa^2 t^2 + \frac{1 - \left(1 - \kappa \right)^{d-1} \left[ 1 + (d-1)\, \kappa \right]}{d \, (d+1)} \, \mathfrak{R}^{2} \,,
\end{align}
where $\mathcal{V}_{d-1} = \mathfrak{R}^{d-1}\,\Omega_{d-2}/ (d-1)$ is
the volume of the sphere (see Ref.~\cite{vish} for an evaluation of galileon Hamiltonians). The first term is the kinetic
energy of a free theory and the second is proportional to the
action polynomial in $(d-1)$ dimensions, i.e.~in one dimension lower. Evaluating the energy on-shell,
we have $E_{\kappa=0} = 0$ for the trivial vacuum and
\begin{equation}
  E_{\kappa=1} = \left( \frac{1}{2} \, t^2 + \frac{1}{d (d+1) } \, \mathfrak{R}^2 \right)\,\mathcal{V}_{d-1}\,\Lambda^{2 \sigma}
 \end{equation}
 for the one at $\kappa=1$. The $\kappa=1$ vacuum can therefore be considered as an excited state, whose energy scales with the cut-off $\Lambda$. 

%%%

\subsection{The galileon duality as a stereographic projection}

\begin{figure}[t!]
\centering
\includegraphics[scale=0.5]{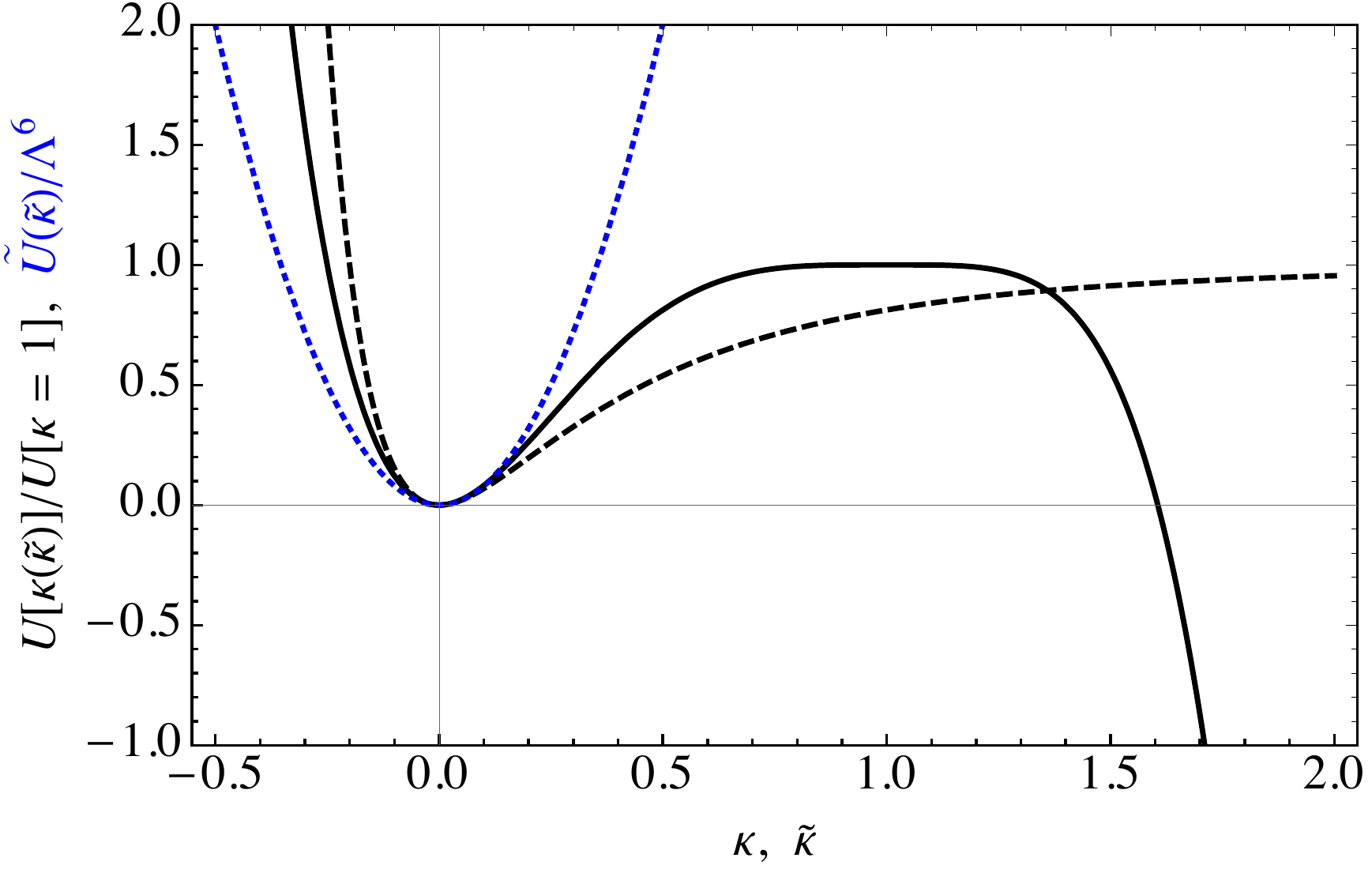}
\caption{\label{fig:pseudopot} Plot of the normalized $\pi$-frame
 action polynomial $U(\kappa)$ as a function of the $\pi$-frame parameter
  $\kappa$ (solid line) and the $\rho$-frame variable
  $\tilde{\kappa}=\kappa/(1-\kappa)$ (dashed line) in $d=4$. The blue dotted line
  depicts the $\rho$-frame action polynomial $\tilde U(\tilde \kappa)/\Lambda^6$.}
\end{figure}

\begin{figure}[t!]
\centering
\includegraphics[scale=0.5]{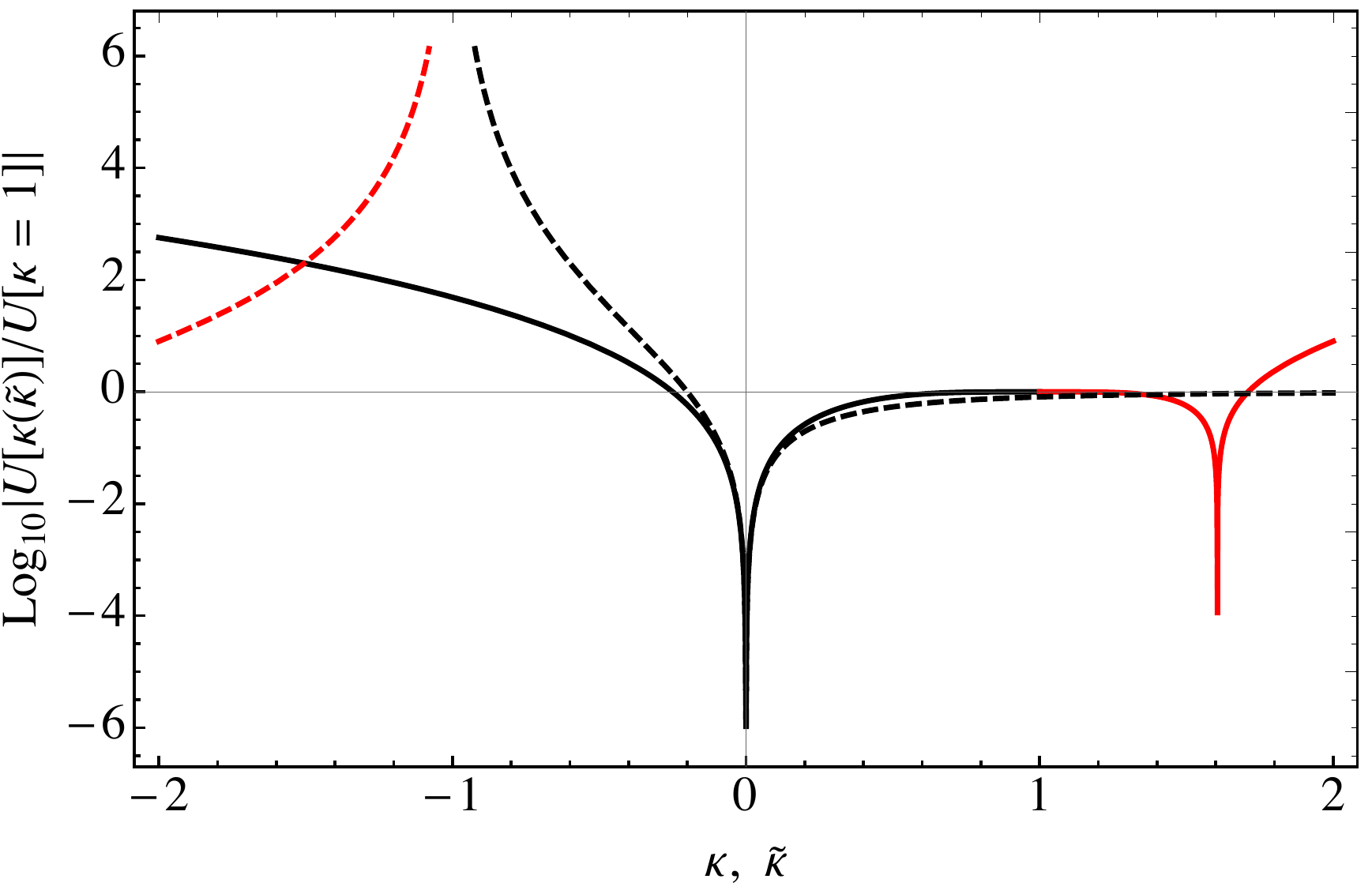}
\caption{\label{fig:pseudopotlog} Logarithmic plot of the normalized action polynomial as a function of the $\pi$-frame parameter $\kappa$ (solid line) and the $\rho$-frame variable $\tilde{\kappa}=\kappa/(1-\kappa)$ (dashed line) in $d=4$. Black lines indicate the region preserved by the duality transformation, and red lines indicate those regions mapped to the left of an infinite barrier in the $\rho$ frame.}
\end{figure}

\begin{figure}[t!]
\centering
\includegraphics[scale=0.16]{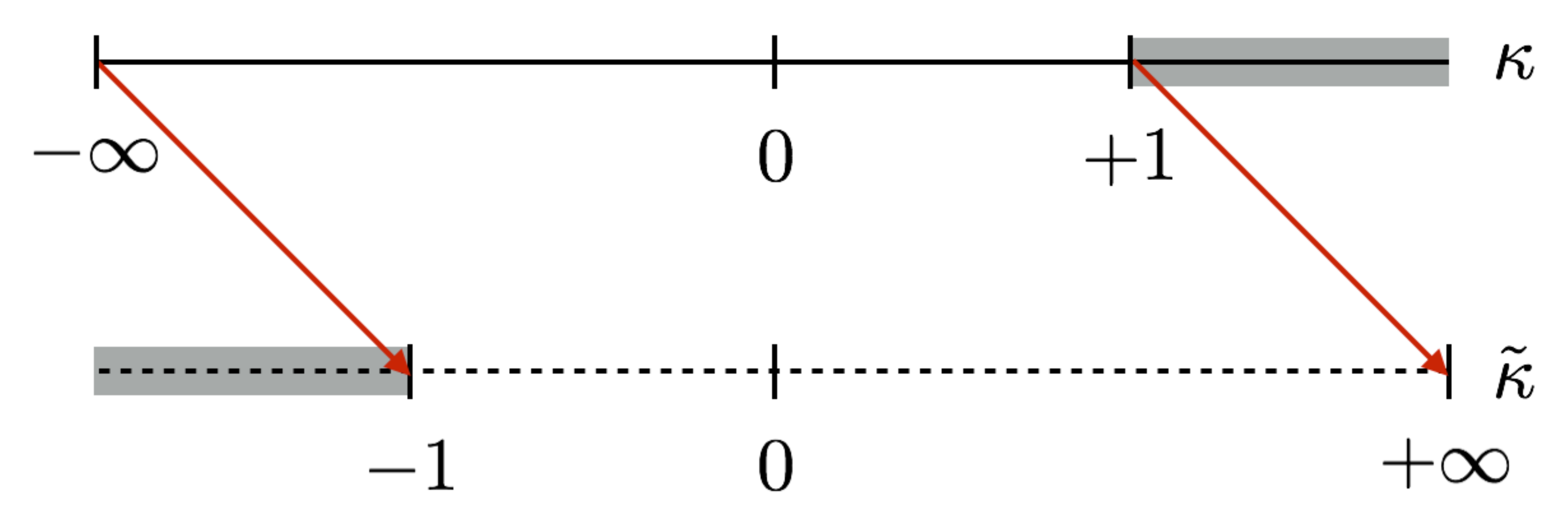}
\caption{\label{fig:mobius} Schematic representation of the M\"{o}bius
  transformation $\tilde{\kappa}=\kappa/(1-\kappa)$, illustrating that
  the $\kappa=1$ vacuum is mapped to positive infinity in the
  $\rho$ frame. The abyssal region (grey shading) is mapped from right
  to left, placing it behind the infinite barrier at
  $\tilde{\kappa}=-1$ (see Figs.~\ref{fig:pseudopot} and
  \ref{fig:pseudopotlog}).}
\end{figure}

We are now interested in how the $\pi$-frame action polynomial is
deformed under the duality transformation. From Eq.~\eqref{eq:duality},
we obtain, in Euclidean signature, 
\begin{align}
  \label{eq:18}
  \pi_\kappa(r) \ \longrightarrow\ \rho_{\tilde \kappa}(\tilde r) \ =\   -\,\frac{\tilde \kappa}{2}\, \Lambda^\sigma \, \tilde r^2\; ,
\end{align}
where we have defined $\tilde \kappa = \kappa / (1-\kappa)$ and  \smash{$\tilde r\ =\ r\: +\: \frac{1}{\Lambda^\sigma} \, \frac{ \rd \pi(r)}{ \rd r}$}. We see
that the $\kappa=1$ vacuum is mapped to infinity and is therefore
removed from the solution space in the $\rho$ frame. This is
visualized in Fig.~\ref{fig:pseudopot}, which depicts the
action polynomial as a function of $\kappa$ (solid line) and
$\tilde \kappa$ (dashed line) for $d=4$.  
The profile for $U(\kappa)$ clearly indicates two stationary points: a trivial stable one at $\kappa=0$ and a strongly-coupled ghost-like vacuum at $\kappa=1$. For an even number of dimensions, off-shell configurations include an abyssal region beyond $\kappa=1$. In contrast, as a function of $\tilde \kappa$,  the
action polynomial $U(\tilde \kappa / (1 + \tilde \kappa))$ only has a
single stationary point at $\tilde \kappa =0$, corresponding to the
trivial vacuum, and seemingly no abyssal region. The transformation $\kappa \to \tilde \kappa$ is a simple example of
the stereographic projection implemented by a M\"obius transformation,
with $\kappa =1$ being the projection point. As a consequence,
the abyssal region for values $\kappa \in (1,\infty)$ is not lost in
the $\rho$ frame. Rather, it is mapped to values
$\tilde \kappa \in (-\infty , -1)$, which reside behind an infinite
``barrier'', as demonstrated in~Fig.~\ref{fig:pseudopotlog}. 
A more schematic
representation of the transformation is provided in
Fig.~\ref{fig:mobius}.

 It is very important to realise that $U(\tilde \kappa / (1 + \tilde \kappa))$ is \emph{not} the action polynomial $\tilde U(\tilde \kappa)$ derived in the $\rho$ frame for a free theory.  The latter can be calculated very easily along the lines described above and is given by 
\begin{align}
  \label{eq:24}
  \tilde U(\tilde \kappa)\  =\   \frac{1}{2} \, d^2  \Lambda^{2 \sigma} \, \tilde \kappa^2 \;.
\end{align}
This is also depicted in Fig.~\ref{fig:pseudopot} (dotted line) and differs decisively from
$U$ as a function of $\tilde \kappa$. In particular, its second
derivative is always positive, suggesting that none of the off-shell
configurations support ghost instabilities. We will see that this
seeming triviality of the $\rho$-frame  polynomial~$\tilde U$ (compared
to $U$) represents a potential pitfall when inferring the stability of
the theory in the presence of external sources (which can support the
off-shell configurations).

%%%

\section{1PI effective action}
\label{sec:effective_action}

A powerful tool for studying non-perturbative aspects of field theories is the effective action. In this section, we will evaluate the 1PI effective action of the model in Eq.~\eqref{eq:pi_action} at order $\hbar$. Taking into account only one saddle point of the classical action at a time, viz.~expanding around one of the two kappa vacua, we will show that the effective action obtains a spurious imaginary part away from the $\kappa=0$ vacuum that is symptomatic of the non-trivial vacuum structure of this galileon theory. Such an imaginary part can often be associated with the decay rate of an unstable state (see Ref.~\cite{Weinberg:1987vp}) that corresponds to some non-perturbative semi-classical solution of the theory.

Before considering the galileon theory directly, it is instructive to work first with a more general theory of a real scalar field $\Phi$. Working in Euclidean signature (where the path integral is well defined), the generating functional of disconnected $n$-point functions has the familiar form
\begin{equation}
\mathcal{Z}[J]\ =\ \int\!\mathcal{D}\Phi\;\exp\bigg[-\:\frac{1}{\hbar}\;\Big(S_{{E}}[\Phi]\:-\:J_x\Phi_x\Big)\bigg]\;,
\end{equation}
where $J_x\equiv J(x)$ is an external source and $S_E=-\,iS$ is the Euclidean action. Throughout, we employ the DeWitt notation in which repeated continuous indices are integrated over, i.e.
\begin{equation}
J_x\Phi_x\ \equiv\ \int\!{\rm d}^dx\;J(x)\Phi(x)\;.
\end{equation}
The generating functional of \textit{connected} $n$-point functions is
\begin{equation}
\mathcal{W}[J]\ =\ -\:\hbar\,\ln\mathcal{Z}[J]\;.
\end{equation}
Taking functional derivatives with respect to the external source, it follows that
\begin{equation}
\frac{\delta^2\mathcal{W}[J]}{\delta J_x\,\delta J_y}\ =\ -\:\frac{1}{\hbar}\Big[\braket{\Phi_x\,\Phi_y}_J\:-\:\braket{\Phi_x}_J\braket{\Phi_y}_J\Big]\;.
\end{equation}
The content of the brackets on the right-hand side is the variance of a distribution, i.e.~a positive semi-definite quantity, and we conclude therefore that
\begin{equation}
\label{eq:concave}
\frac{\delta^2\mathcal{W}[J]}{\delta J_x\,\delta J_y}\ <\ 0\;,
\end{equation}
that is, $\mathcal{W}[J]$ is a concave functional of the source $J$.

The one-particle irreducible effective action~\cite{Jackiw1974} is defined via the Legendre transformation
\begin{equation}
\Gamma[\phi]\ =\ \max_{J}\Big[\mathcal{W}[J]\:+\:J_x\phi_x\Big]\;,
\end{equation}
which, given Eq.~\eqref{eq:concave}, is a convex functional of $\phi$. After performing the extremization, the external source $J_x$ is a fixed functional of $\phi$, i.e.~$J_x\to\mathcal{J}_x\equiv\mathcal{J}_x[\phi]$, and we have
\begin{equation}
\phi_x\ \equiv\ \braket{\Phi_x}_{\mathcal{J}}\ =\ -\:\frac{\delta \mathcal{W}[\mathcal{J}]}{\delta \mathcal{J}_x}\;,
\end{equation}
with the 1PI effective action taking the form
\begin{equation}
\Gamma[\phi]\ =\ \mathcal{W}[\mathcal{J}]\:+\:\mathcal{J}_x[\phi_x]\phi_x\;.
\end{equation}
The 1PI effective action yields the \textit{source-dependent} quantum equation of motion
\begin{equation}
\frac{\delta \Gamma[\phi]}{\delta \phi_x}\ =\ \mathcal{J}_x[\phi]\;.
\end{equation}
When evaluated at the extremal field configuration, which we denote by $\varphi$, this yields the more familiar expression
\begin{equation}
\frac{\delta \Gamma[\phi]}{\delta \phi}\bigg|_{\phi\,=\,\varphi}\ =\ \mathcal{J}_x[\varphi]\ =\ 0\;.
\end{equation}
Note that the source vanishes at the extremal field configuration, but it is non-zero for classically off-shell field configurations. We will see this explicitly later. For further details of this point, see Ref.~\cite{Garbrecht2016}.

It is known that perturbative evaluations of the effective action can violate the above-mentioned convexity, and such non-convexity is a signal of non-trivial vacuum structure or instabilities of the theory. An archetypal example is the scalar theory with a non-convex classical potential
\begin{equation}
U(\Phi)\ =\ -\:\frac{1}{2!}\,\mu^2\,\Phi^2\:+\:\frac{1}{4!}\,\lambda\,\Phi^4\;.
\end{equation}
If we evaluate the effective action of this theory by expanding perturbatively around only one vacuum, assuming a homogeneous background field configuration $ \braket{\Phi_x} = \phi = \text{const}$, the effective potential has a non-convex region where it obtains a non-zero imaginary part. Instead, if we were to sum over all saddle points of the classical action (see e.g.~Ref.~\cite{Plascencia2016a}), we would find that the effective potential is convex, with the Maxwell construction arising as a result of the homogeneous superposition of both vacuum states (see e.g.~Ref.~\cite{Alexandre1999,Alexandre2013}).

In what follows, we will show that the 1PI effective action for the galileon theory in Eq.~\eqref{eq:pi_action} exhibits such a violation of convexity, indicating non-trivial vacuum structure. The 1PI effective action has the form
\begin{equation}
\Gamma[\uppi]\ =\ \mathcal{W}[\mathcal{J}]\:+\:\mathcal{J}_x[\uppi]\uppi_x\;,
\end{equation}
where
\begin{equation}
\mathcal{W}[\mathcal{J}]\ =\ -\:\hbar\,\ln \mathcal{Z}[\mathcal{J}]\;,\qquad \mathcal{Z}[\mathcal{J}]\ =\ \int\!\mathcal{D}\pi\;\exp\bigg[-\:\frac{1}{\hbar}\,\Big(S_{E}[\pi]\:-\:\mathcal{J}_x\pi_x\Big)\bigg]\;,
\end{equation}
and
\begin{equation}
\uppi_x\ =\ -\:\frac{\delta \mathcal{W}[\mathcal{J}]}{\delta \mathcal{J}_x}\;.
\end{equation}

It will prove illustrative to evaluate the effective action by means of the duality transformation. Up to a choice of normalization, the same result is obtained by evaluating the effective action directly in terms of the $\pi$ variable. This will illustrate the importance of the source in keeping track of both vacua. In order to apply the duality transformation at the level of the path integral, we insert unity in the form
\begin{equation}
  1\ =\ \int\mathcal{D}\rho\;\delta(\pi-\pi[\rho])\,\bigg|\mathrm{det}\,\frac{\delta \pi[\rho]}{\delta \rho}\bigg|\;.
\end{equation}
In order to avoid proliferation of symbolic variants of $\pi$, we disambiguate the explicit configuration $\pi[\rho]$ from the functional variable $\pi$ only by the presence of the functional argument. 
The functional determinant is unity (see Ref.~\cite{Kampf2014}), and we obtain
\begin{equation}
\mathcal{Z}[\mathcal{J}]\ =\ \int\!\mathcal{D}\rho\;\exp\bigg[-\:\frac{1}{\hbar}\Big(\tilde{S_E}[\rho]\:-\:\mathcal{J}_x[\uppi]\,\pi_x[\rho]\Big)\bigg]\;.
\end{equation}
This expression merits further comment. Although the dual action $\tilde{S}_E[\rho]$ does indeed correspond to a trivial theory, there is a highly non-trivial coupling between the dual field variable $\rho$ and the source. As observed in Ref.~\cite{Kampf2014}, $\rho_x$ and $\pi_x[\rho]$ share the same one particle poles at $p^2=0$, up to a rescaling of the residues.  This guarantees the equivalence of on-shell  S-matrices, even at the loop level, provided we treat the system perturbatively about the trivial vacuum, which is present in both frames. However, our interest here, and indeed in Ref.~\cite{Keltner2015}, lies in the deep UV behaviour of the theory above and beyond the scale of strong coupling, when perturbative unitarity breaks down and the non-trivial vacuum structure of the $\pi$ frame begins to emerge. As we will see, the source coupling in the $\rho$ frame allows us to keep track of that non-trivial  vacuum structure.

To this end, we now perform a saddle-point evaluation of the path integral, expanding $\rho$ around the solution $\varrho$ of
\begin{equation}
\label{eq:stationarity}
\frac{\delta \tilde{S}_E[\rho]}{\delta \rho_{\tilde{x}}}\bigg|_{\rho\,=\,\varrho}\ = \ \mathcal{J}_y[\uppi]\,\frac{\delta \pi_y[\rho]}{\delta \rho_{\tilde{x}}}\bigg|_{\rho\,=\,\varrho}\;,
\end{equation}
writing $\rho =\varrho\:+\:\hbar^{1/2}\hat{\rho}$. Note that there is an implicit integral over $y$ on the right-hand side of Eq.~\eqref{eq:stationarity}. Expanding to order $\hbar$, we have
\begin{equation}
\mathcal{Z}[\mathcal{J}]\ =\ \exp\bigg[-\:\frac{1}{\hbar}\Big(\tilde{S}_E[\varrho]\:-\:\mathcal{J}_x[\uppi]\,\pi_x[\varrho]\Big)\bigg]\int\!\mathcal{D}\hat{\rho}\;\exp\bigg[-\:\frac{1}{2}\,\hat{\rho}_{\tilde{x}}\,G_{\tilde{x}\tilde{y}}^{-1}(\mathcal{J},\varrho)\,\hat{\rho}_{\tilde{y}}\:+\:\dots\bigg]\;,
\end{equation}
where we have defined
\begin{equation}
  \label{eq:Ginrhoframe}
G^{-1}_{\tilde{x}\tilde{y}}(\mathcal{J},\varrho)\ \equiv\ \frac{\delta^2 \tilde{S}_E[\rho]}{\delta \rho_{\tilde{x}}\delta\rho_{\tilde{y}}}\bigg|_{\rho\,=\,\varrho}\:-\:\mathcal{J}_z[\uppi]\,\bigg[\frac{\delta^2\pi_z[\rho]}{\delta \rho_{\tilde{x}}\delta \rho_{\tilde{y}}}\bigg]_{\rho\,=\,\varrho}\;.
\end{equation}
We see that the non-triviality of the original theory has been maintained in the source-dependent term. Naively, we might expect that the source-dependent term in Eq. \eqref{eq:Ginrhoframe} vanishes for on-shell configurations and we are left with a trivial free-theory propagator. This is indeed true for the trivial vacuum, $\kappa=0$. However, as we approach the non-trivial vacuum,  $\kappa \to 1$,  we will see that the divergence in  the duality map in this limit compensates for the vanishing of the source, and there remains a residual contribution. The resulting effective action obtains a spurious imaginary part.

To see this, we perform the integral over the quadratic fluctuations to obtain the following effective action\footnote{Of course, in the presence of non-convex regions, careful treatment of the non-Gaussian functional integral is required, for instance by appropriate analytic continuation (see e.g.~Ref.~\cite{Andreassen:2016cvx}). For our purposes, however, the perturbative analysis suffices to diagnose non-trivial behaviour of the theory.}
\begin{equation}
\label{eq:effact1}
\Gamma[\uppi]\ =\ \tilde{S}_E[\varrho]\:+\:\mathcal{J}_x[\uppi](\uppi_x-\pi_x[\varrho])\:+\:\frac{\hbar}{2}\,\mathrm{tr}\,\ln\,G^{-1}(\mathcal{J},\varrho)\ast G(0,0)\:+\:\mathcal{O}(\hbar^2)\;,
\end{equation}
where $\ast$ denotes a convolution and $G_{xy}(0,0)$ appears for normalization. We can now proceed by eliminating $\pi$ and $\mathcal{J}$ in favour of $\uppi$. Since $\tilde{S}_E[\varrho]=S_E[\pi]$ ($\pi\equiv\pi[\varrho]$) and $\uppi-\pi=\mathcal{O}(\hbar)$ (formally), we write
\begin{equation}
S_E[\pi]\ =\ S_E[\uppi]\:-\:\frac{\delta S_E[\uppi]}{\delta \uppi_x}\bigg|_{\uppi\, =\, \pi}\big(\uppi_x-\pi_x[\varrho]\big)\:+\:\mathcal{O}(\hbar^2)
\end{equation}
and use the stationarity condition
\begin{equation}
\label{eq:pisaddle}
\frac{\delta S_E[\uppi]}{\delta \uppi_x}\bigg|_{\pi_x}\ =\ \mathcal{J}_x[\uppi]
\end{equation}
to eliminate the second term on the right-hand side of Eq.~\eqref{eq:effact1}. Note that Eq.~\eqref{eq:pisaddle} is consistent with Eq.~\eqref{eq:stationarity}. At order $\hbar$, we then have
\begin{equation}
\label{eq:effact2}
\Gamma[\uppi]\ =\ S_E[\uppi]\:+\:\frac{\hbar}{2}\,\mathrm{tr}\,\ln\,G^{-1}(\mathcal{J},\varrho)\ast G(0,0)\;.
\end{equation}
If we choose to perform the saddle-point evaluation of the path integral around the maximally symmetric solutions $\pi_{\kappa}(x)$ and by comparing Eqs.~\eqref{eq:eom_pi} and \eqref{eq:pisaddle}, we see that the explicit form of the source is
\begin{equation}
\label{eq:offshellJ}
\mathcal{J}_x[\uppi]\ =\ d\,\kappa(1-\kappa)^{d-1}\,\Lambda^{\sigma}\;,
\end{equation}
vanishing on-shell, i.e.~when $\kappa=0$ or $1$, as we would expect. We shall now show,  however, that the source-dependent term
\begin{equation}
\mathcal{J}_z[\uppi]\,\bigg[\frac{\delta^2\pi_z[\rho]}{\delta \rho_{\tilde{x}}\delta \rho_{\tilde{y}}}\bigg]_{\rho\,=\,\varrho}
\end{equation}
in Eq.~\eqref{eq:Ginrhoframe} does not vanish in the limit $\kappa\to 1$.

Since we know from Eq.~\eqref{eq:offshellJ} that $\mathcal{J}_x[\uppi]$ is a constant for off-shell, maximally symmetric solutions, we need to evaluate
\begin{equation}
\int\!{\rm d}^dz\;\bigg[\frac{\delta^2\pi[\rho](z)}{\delta \rho(\tilde{x})\delta \rho(\tilde{y})}\bigg]_{\rho\,=\,\varrho}\ =\ \Bigg[\int\!{\rm d}^dz\;\frac{\delta}{\delta \rho(\tilde{y})}\,\delta^d(z+\partial\pi[\rho](z)/\Lambda^{\sigma}-\tilde{x})\Bigg]_{\rho\,=\,\varrho}\;,
\end{equation}
cf.~Eq.~\eqref{eq:5}. After performing the remaining functional derivative, we have
\begin{align}
&\int\!{\rm d}^dz\;\bigg[\frac{\delta^2\pi[\rho](z)}{\delta \rho(\tilde{x})\delta \rho(\tilde{y})}\bigg]_{\rho\,=\,\varrho}\nonumber\\&\qquad  =\ \Bigg[\int\!{\rm d}^dz\;\frac{\partial}{\partial a_{\mu}}\,\frac{\delta^d(a)}{\Lambda^\sigma}\bigg|_{a\,=\,z+\frac{\partial\pi[\rho](z)}{\Lambda^{\sigma}}-\tilde{x}}\,\frac{\partial}{\partial z_{\mu}}\,\delta^d(z+\partial\pi[\rho](z)/\Lambda^{\sigma}-\tilde{y})\Bigg]_{\rho\,=\,\varrho}\;,
\end{align}
which, when evaluated on the set of maximally symmetric solutions, yields
\begin{equation}
\int\!{\rm d}^dz\;\bigg[\frac{\delta^2\pi[\rho](z)}{\delta \rho(\tilde{x})\delta \rho(\tilde{y})}\bigg]_{\rho\,=\,\varrho}\ =\ \frac{1}{\Lambda^{\sigma}}\int\!{\rm d}^dz\;\frac{\partial}{\partial a_{\mu}}\,\delta^d(a)\bigg|_{a\,=\,z(1-\kappa)-\tilde{x}}\,\frac{\partial}{\partial z_{\mu}}\,\delta^d(z(1-\kappa)-\tilde{y})\;.
\end{equation}
This can be re-written in the form
\begin{equation}
\int\!{\rm d}^dz\;\bigg[\frac{\delta^2\pi[\rho](z)}{\delta \rho(\tilde{x})\delta \rho(\tilde{y})}\bigg]_{\rho\,=\,\varrho}\ =\ \frac{1}{\Lambda^{\sigma}(1-\kappa)}\int\!{\rm d}^dz\;\frac{\partial}{\partial z_{\mu}}\,\delta^d(z(1-\kappa)-\tilde{x})\,\frac{\partial}{\partial z_{\mu}}\,\delta^d(z(1-\kappa)-\tilde{y})\;.
\end{equation}
Making the change of variables $\tilde{z}=(1-\kappa)z$ and integrating by parts, we can show that
\begin{equation}
\int\!{\rm d}^dz\;\bigg[\frac{\delta^2\pi[\rho](z)}{\delta \rho(\tilde{x})\delta \rho(\tilde{y})}\bigg]_{\rho\,=\,\varrho}\ =\ -\:\frac{1}{\Lambda^{\sigma}(1-\kappa)^{d-1}}\,\tilde{\Delta}\delta^d_{\tilde{x}\tilde{y}}\;.
\end{equation}
As anticipated, this is singular as $\kappa \to 1$, with the degree of divergence exactly compensating for the vanishing of the source in the same limit, as per Eq. \eqref{eq:offshellJ}. The corresponding poles and  zeroes cancel, and we are left with
\begin{equation}
\label{eq:Ginrhoframe2}
G^{-1}_{\tilde{x}\tilde{y}}(\mathcal{J},\varrho)\ \equiv\ \frac{\delta^2 \tilde{S}_E[\rho]}{\delta \rho_{\tilde{x}}\delta\rho_{\tilde{y}}}\bigg|_{\rho\,=\,\varrho}\:-\:\mathcal{J}_z[\uppi]\,\bigg[\frac{\delta^2\pi_z[\rho]}{\delta \rho_{\tilde{x}}\delta \rho_{\tilde{y}}}\bigg]_{\rho\,=\,\varrho}\ =\ -\:(1-d\kappa)\tilde{\Delta}\delta^d_{\tilde{x}\tilde{y}}\;.
\end{equation}
We might worry that this differs from [see Eq.~\eqref{eq:2nd_var}]
\begin{equation}
G^{-1}_{x y}(\mathcal{J},\uppi)\ \equiv\ \frac{\delta^2 {S}_E[\pi]}{\delta \pi_x \delta \pi_y}\bigg|_{\pi\,=\,\uppi}\ =\ -\:(1-\kappa)^{d-2}(1-d\kappa ) \Delta \delta^d_{xy}\;,
\end{equation}
as obtained directly in terms of $\pi$ variables, by a factor of $(1-\kappa)^{d-2}$. However, returning to the original functional integral over the quadratic fluctuations about the kappa configurations, we have
\begin{equation}
\int\!{\rm d}^d\tilde{x}\,{\rm d}^d\tilde{y}\;\hat{\rho}(\tilde{x})(1-d\kappa)[\tilde{\Delta}\delta^d(\tilde{x}-\tilde{y})]\hat{\rho}(\tilde{y})\ =\ \int\!{\rm d}^dx\,{\rm d}^dy\;\hat{\pi}(x)(1-\kappa)^{d-2}(1-d\kappa)[\Delta\delta^d(x-y)]\hat{\pi}(y)\;,
\end{equation}
where $\hat{\pi}(x)=\hat{\rho}(\tilde{x})$. We see, then, that the calculation of the 1PI effective action is frame independent, as it should be, so long as we choose the correct normalization and keep track of all the appropriate source-dependent terms. Therefore, choosing a consistent normalization relative to the $\kappa=0$ vacuum, we arrive at the following result for the 1PI effective action at order $\hbar$:
\begin{equation}
\Gamma[\uppi]\ =\ S_E[\uppi]\:+\:\frac{\hbar}{2}\,\Omega\,\ln\Big[(1-\kappa)^{d-2}(1-d\kappa)\Big]\;,
\end{equation}
where $\Omega$ is a real-valued phase-space volume factor that comprises a $d$-dimensional coordinate-space integral and a $d$-dimensional momentum-space integral.\footnote{The phase-space volume factor $\Omega$ depends on the choice of regularization and renormalization scheme, being non-vanishing and real in dimensional regularization. Its precise value, however, is unimportant for our discussions, and the effective action serves only as a powerful diagnostic tool, wherein a finite (or infinite) imaginary part is indicative of non-trivial vacuum structure.} For $\kappa = 0$ or $d=1$, we therefore have
\begin{equation}
\Gamma[\uppi]\ =\ S_E[\uppi]\:+\:\mathcal{O}(\hbar^2)\;.
\end{equation}
This result is consistent with the observation that there are no one-loop corrections about the $\kappa=0$ vacuum~\cite{kurtmgal,Kampf2014,Goon2016}. For $\kappa=1$ and $d\geq 2$, the effective action is ill-defined at order $\hbar$, and this is indicating that we must work to higher order, i.e.~proceeding via the 4PI effective action (see e.g.~Ref.~\cite{Carrington:2004sn}), so as to resum conveniently the non-trivial behaviour that is arising at fourth order. For $\kappa<1/d$, the one-loop 1PI effective action remains real. Instead, for $\kappa>1/d$, the one-loop 1PI effective action obtains a spurious imaginary part that is symptomatic of the violation of convexity:
\begin{equation}
\Gamma[\uppi]\ =\ S_E[\uppi]\:+\:\frac{\hbar}{2}\,\Omega\,\ln\Big|(1-\kappa)^{d-2}(1-d\kappa)\Big|\:+\:\frac{i\pi\hbar}{2}\,\Omega
\begin{cases}1\;,&1/d\ <\ \kappa\ <\ 1\;,\ d\ >\ 1\;,\\
1\;,&\kappa\ >\ 1\;,\ d\ >\ 1\;,\ d\text{\ even}\;,\\
0\;,&\text{otherwise}\;,
\end{cases}
\end{equation}
where we have restricted to the principal branch of the logarithm. This observed violation of convexity should be anticipated from our analysis of the action polynomial, and we certainly should not trust our naive calculation of the effective action in the non-convex region. The result, however, is useful for diagnosing (tachyonic or ghost) instabilities and/or signalling the possibility of non-perturbative semi-classical contributions to the path integral, arising for instance from instantons or sphalerons. In fact, one might be tempted to interpret the $\kappa=1$ solution --- sitting at a maximum of the action polynomial between the local minimum and the abyssal region (for even $d$) --- as such a sphaleron. (For discussions of solitons in the context of galileon theories, see Ref.~\cite{Carrillo-Gonzalez:2016yxq} and references therein.) Nevertheless, this aspect requires significant further study beyond the scope of the present article, and we will not comment on it further. The most important observation is that any such solutions (or vacua) will become relevant at scales of order $\Lambda$, and, in the next section, we will discuss the implications of this for the spectral density of the theory.

%%%

\section{Wightman functions}

One important lesson from the effective-action analysis of the preceding section is that we
have to include external sources in order to keep track of the
non-trivial vacuum structure of the theory when we employ the
duality. Here, we demonstrate how this affects the calculation of the
Wightman functions, {studied} in Ref.~\cite{Keltner2015}.

Our starting point is the Euclidean Wightman function in the presence
of an external source $\mathcal{J}$:
\begin{align}
\braket{\pi_x \, \pi_y}_{\mathcal{J}}\ =\ \frac{1}{\mathcal{Z}[\mathcal{J}]} \int \mathcal{D} \pi \, \pi_x \, \pi_y \exp \left[ - \frac{1}{\hbar} \left( S_E[\pi] - \mathcal{J}_z \, \pi_z \right)\right]\;,
\end{align}
where $x^0 \geq y^0$. In the $\pi$ frame, this object can only be calculated perturbatively
without referring to a particular UV completion. In Ref.~\cite{Keltner2015}, it was therefore suggested to use the duality
to express the right-hand side entirely in terms of $\rho$. After all, the
\textit{source-free} $\rho$-frame theory is UV complete in a trivial
way. The corresponding path integral expression in the $\rho$ frame
reads
\begin{align}
  \label{eq:rho_wightman}
\braket{\pi_x \, \pi_y}_{\mathcal{J}}\ =\ \frac{1}{\mathcal{Z}[\mathcal{J}]} \int \mathcal{D} \rho \, \pi_x[\rho] \, \pi_y[\rho] \exp \left[ - \frac{1}{\hbar} \left( \tilde S_E[\rho] - \mathcal{J}_z \, \pi_z[\rho] \right)\right]\;.
\end{align}
As before, we perform a saddle-point evaluation of the path integral,
expanding around solutions $\varrho$ of the classical equation of
motion in Eq.~\eqref{eq:stationarity}. At order $\hbar$, we find
\begin{align}
  \braket{\pi_x \, \pi_y}_{\mathcal{J}}\  &=\ \frac{1}{\mathcal{Z}[\mathcal{J}]} \sum_\mathrm{saddles} \exp\bigg[- \frac{1}{\hbar}\Big(\tilde{S}_E[\varrho]\:-\:\mathcal{J}_z\,\pi_z[\varrho]\Big)\bigg]\nonumber\\&\qquad \times\:  
  \int\!\mathcal{D}\hat{\rho} \; \pi_x[\rho] \, \pi_y[\rho] \exp\bigg[-\frac{1}{2}\,\hat{\rho}_{\tilde{x}}\,G_{\tilde{x}\tilde{y}}^{-1}(\mathcal{J},\varrho)\,\hat{\rho}_{\tilde{y}}\:+\:\dots\bigg]\;,
\end{align}
where $G_{\tilde{x}\tilde{y}}^{-1}(\mathcal{J},\varrho)$ was defined
in Eq.~\eqref{eq:Ginrhoframe}. When evaluated on the set of maximally symmetric solutions we have, according to Eq.~\eqref{eq:Ginrhoframe2},
that
\begin{equation}
  \label{eq:GwithJ}
G_{\tilde{x}\tilde{y}}^{-1}(\mathcal{J},\varrho)\ =\ - \left[1 - \frac{{\mathcal{J}}_\kappa}{\left(1-\kappa\right)^{d-1}{\Lambda^{\sigma}}}\right]\tilde{\Delta}\,\delta^d_{\tilde{x}\tilde{y}}\;,
\end{equation}
where the source $\mathcal{J}_\kappa$ is given in
Eq.~\eqref{eq:offshellJ} and supports the corresponding kappa
configuration. We recall that $ \mathcal{J}_\kappa $ vanishes for the
on-shell configurations $\kappa = 0$ and $\kappa =1$, as one would
expect. {Nevertheless}, both saddle points {must} contribute to
$\lim_{\mathcal{J} \to 0} \braket{\pi_x \,\pi_y}_{\mathcal{J}}$ {in both frames}.

In the case of the trivial vacuum ($\kappa =0$),
$ \tilde S_E[\varrho] =0 $, and the relevant contribution reads
\begin{equation}
  \label{eq:free_wight}
 \lim_{\mathcal{J} \to 0} \braket{\pi_x \,\pi_y}_{\mathcal{J}}\ \supset\ \frac{1}{\mathcal{Z}[{0}]}  \int\!\mathcal{D}\hat{\rho} \; \pi_x[\hat \rho] \, \pi_y[\hat \rho] \exp\bigg[ \frac{1}{2}\, \hat{\rho}_{\tilde{x}}\, \tilde \Delta \hat{\rho}_{\tilde{x}}\bigg]\;,
\end{equation}
which matches the free-theory expression evaluated in Ref.~\cite{Keltner2015} (after rotating back to real time). Since all
higher-order terms in the expansion vanish, it is tempting to trust it
for energies above the strong-coupling scale~$\Lambda$. However, that
reasoning ignores the presence of at least one further highly
non-trivial vacuum\footnote{We have restricted the discussion to the
  subspace of maximally symmetric configurations. It is possible
  though that further saddle points have been mapped out by the
  duality.}. To be specific, in the limit $\kappa \to 1$, the second
term in square brackets in Eq.~\eqref{eq:GwithJ} gives a finite
contribution (despite the vanishing of $\mathcal{J}_\kappa$), which
changes the sign of the kinetic operator for $d>1$. Correspondingly,
this second saddle point contributes
\begin{equation}
  \label{eq:interact_wight}
 \lim_{\mathcal{J} \to 0} \braket{\pi_x \,\pi_y}_{\mathcal{J}}\ \supset\  \frac{1}{\mathcal{Z}[{0}]} \, \exp\left[ -\frac{1}{\hbar} \, \tilde S_E[\varrho] \right] \int\!\mathcal{D}\hat{\rho} \; \pi_x[\hat \rho] \, \pi_y[\hat \rho] \exp\bigg[-\frac{d-1}{2}\,\hat{\rho}_{\tilde{x}}\,\tilde \Delta \,\hat{\rho}_{\tilde{x}} \, + \, \ldots\bigg]\;.
\end{equation}
Apart from the pathological sign, which signals a ghost instability
around that vacuum (in accordance with the effective-action
calculation), the higher-order terms in the saddle-point expansion
(indicated by the dots and suppressed by additional powers of
  $\Lambda^\sigma$) are generically non-vanishing. In fact, they
re-introduce the complexity of the original $\pi$-frame
expression. This can be seen by taking higher functional
$\rho$-derivatives of $\mathcal{J}_z \, \pi_z[\rho]$ in
Eq.~\eqref{eq:rho_wightman}.

We therefore conclude that the free-theory expression in
Eq.~\eqref{eq:free_wight} is not sufficient to infer the high-energy
behavior of the Wightman functions. Rather, we have to take into
account the non-trivial vacuum structure, which, at least in principle,
can be preserved by including external sources, but is lost by naive
application of the duality and careless treatment of the limit
$\mathcal{J} \to 0$. 

%%%

\section{Conclusion}

Due to the non-renormalizable nature of their interactions, galileons are only properly understood as effective field theories with a cut-off.  Further, to be phenomenologically interesting, galileon interactions must operate on macroscopic scales, forcing the cut-off to be unacceptably low and elimininating the predictive power even at the scale of experimental probes. This demands a better understanding of the UV properties of galileons, a task that has been pioneered by Keltner and Tolley~\cite{Keltner2015}. In a very thoughtful piece of work, they have argued that galileons fall into a class of non-localizable field theories, whose UV completion should be non-Wilsonian, possibly along the lines of classicalization~\cite{Dvali2011a}.

In this paper, we have studied galileon duality from a quantum perspective and asked what inferences can really be drawn regarding the UV sector of galileon theories.   Mirroring the classical analysis of Ref.~\cite{Creminelli2015}, we emphasize the importance of   keeping proper track of the source dependence. In doing so, we reveal that the calculation of spectral densities on the ``free'' theory side is sensitive to a non-trivial vacuum structure that is easily missed if source terms are  not carefully tracked.  The non-trivial vacuum structure can, of course, be anticipated by considering the original interacting galileon theory.  This has two ``kappa vacua'': one with $\kappa=0$ and a second with $\kappa=1$.  It turns out that the galileon duality acts as a simple stereographic projection
  on the space of  these ``kappa configurations'', with the $\kappa=1$
  solution being the projection point that is \emph{mapped out of the
  physical configuration space}. The best way to avoid this is to couple $\pi$
  to an external source, thereby maintaining the vacuum structure of the original theory.

When we perform the effective action calculation on the ``free'' theory side, it turns out that the $\kappa \to 1$ limit is highly non-trivial. Although the source vanishes in this limit, as of course it should for an on-shell configuration, the map itself diverges, and the two effects compensate to leave a residual contribution.  A saddle-point evaluation around a generic kappa configuration of the 1PI
  effective action signals a violation of convexity for
  $\kappa > 1/d$ ($d>1$). This is indicative of {non-trivial vacuum structure}, {potentially supporting non-perturbative classical solutions that may contribute} to the {spectral density}.
  
 With a view to asking what  we can really understand about the UV sector of galileon theories,  we have evaluated contributions to the position-space Wightman functions explicitly by expanding about various saddle points.   We can compare this with Ref.~\cite{Keltner2015}, where they  compute the contribution from the trivial saddle point, corresponding to the $\kappa=0$ vacuum, but do not include the contribution from a second saddle, corresponding to the $\kappa=1$ vacuum. The latter contribution is non-trivial and {\it cannot} be neglected. Furthermore, its explicit cut-off dependence significantly complicates a direct
  calculation of the high-energy scaling. The result is that one cannot reliably calculate  the scaling behaviour of the spectral density beyond the cut-off of the low-energy effective theory. As this was the main  piece of concrete evidence in support of the idea that galileons \emph{must} UV complete via a non-Wilsonian mechanism, we conclude that such conclusions are premature. More generally, our analysis suggests that galileon duality is not a useful tool for understanding the UV properties of galileons.

%%%

\begin{acknowledgments}

This work was supported by STFC grant ST/L000393/1. PM would like to
thank Massimo Porrati and the members of the Centre for Particle
Cosmology at the University of Pennsylvania for helpful discussions
during the preparation of this manuscript. FN would like to thank
Stefan Hofmann for inspiring discussions.

\end{acknowledgments}

%%%

\bibliography{galileonduality}{}

\end{document}